\documentclass[a4paper,cite]{article}
\usepackage{latexsym}
\newcommand{\lbl}[1]{\label{eq:#1}}
\newcommand{ \rf}[1]{(\ref{eq:#1})}

\newcommand{\be}{\begin{equation}}
\newcommand{\ee}{\end{equation}}
\newcommand{\bea}{\begin{eqnarray}}
\newcommand{\eea}{\end{eqnarray}}
\newcommand{\setl}{\setlength\arraycolsep{2pt}}

\newcommand{\noi}{\noindent}
\newcommand{\nn}{\nonumber}
\newcommand{\ra}{\rightarrow}
\newcommand{\Ra}{\Rightarrow}

\newcommand{\lesssim}{ {\
\lower-1.2pt\vbox{\hbox{\rlap{$<$}\lower5pt\vbox{\hbox{$\sim$}}}}\ }
}
\newcommand{\gtrsim}{ {\
\lower-1.2pt\vbox{\hbox{\rlap{$>$}\lower5pt\vbox{\hbox{$\sim$}}}}\ }
}

\newcommand{\cA}{{\cal A}}

\newcommand{\cM}{{\cal M}}

\newcommand{\cO}{{\cal O}}

\newcommand{\cR}{{\cal R}}

\newcommand{\Imm}{\mbox{\rm Im}}
\newcommand{\Ree}{\mbox{\rm Re}}

\newcommand{\with}{\mbox{\rm with}}

\newcommand{\annd}{\mbox{\rm and}}
\newcommand{\foor}{\mbox{\rm for}}

\input epsf

\def\theequation{\arabic{section}.\arabic{equation}}

\begin{document}

\begin{titlepage}

\vspace*{1.5cm}
\begin{center}
{\LARGE \bf  Testing an Approximation to Large-$N_c$ QCD \\[0.5cm]
with a Toy Model}\\[1cm]

{\bf Maarten Golterman}$^a$,
 {\bf Santiago Peris}$^b$, {\bf Boris Phily}$^c$  and  {\bf Eduardo de
Rafael}$^c$\\[1cm]

$^a$ Institute for Nuclear Theory, University of Washington\\
Box 351550, Seattle, WA 98195, USA, and\\
Dept. of Physics and Astronomy, San Francisco State University\\
1600 Holloway Ave, San Francisco, CA 94132, USA\\
[0.5cm]
$^b$ Grup de F{\'\i}sica Te{\`o}rica and IFAE\\ Universitat
Aut{\`o}noma
de Barcelona, 08193 Barcelona, Spain\\
[0.5cm]
$^c$  Centre  de Physique Th{\'e}orique\\
       CNRS-Luminy, Case 907\\
    F-13288 Marseille Cedex 9, France\\

\end{center}

\vspace*{1.0cm}

\begin{abstract}
We consider a simple model of large--$N_c$ QCD  defined by a
spectrum  consisting of an infinite set of equally spaced
zero-width vector resonances. This model is an excellent
theoretical laboratory for investigating certain approximation
schemes which have been used recently in calculations of hadronic
parameters, such as the Minimal Hadronic Approximation. We also
comment on some of the questions concerning issues of {\it local
duality} versus {\it global duality} and finite--energy sum rules.

\end{abstract}

\end{titlepage}

\section{\normalsize Introduction}
\lbl{int}

\noi The suggestion to keep the number of colours  $N_c$ in QCD
as a free parameter was made by 't~Hooft~\cite{THFT74} as a
possible way to approach the study of non--perturbative
phenomena. In spite of the efforts of many illustrious theorists
who have worked on the subject, QCD in the large--$N_c$ limit
still remains unsolved. However, many interesting properties have
been proved\footnote{For recent reviews, see e.g. the book in
ref.~\cite{BW93} and the lectures in ref.~\cite{Man99}.} which suggest
that, indeed, the theory in this limit has the bulk of the
non--perturbative properties of three--colour QCD. In particular, it
has been shown that, if confinement persists in this limit, there
is spontaneous chiral symmetry breaking~\cite{CW80}.

The spectrum of the theory in the large--$N_c$ limit consists of
an infinite number of narrow stable meson states which are
flavour nonets~\cite{W79}. This spectrum, however, looks {\it a
priori} not quite like the one in the real world. The vector and
axial--vector spectral functions measured in $e^+ e^- \ra$
hadrons and in hadronic $\tau$ decay show indeed a richer
structure than just a sum of narrow states. There are, however,
many instances where one is only interested in observables which
are given by weighted integrals of hadronic spectral functions.
In these cases, it may be enough to know a few {\it global}
properties of the hadronic spectrum in order to have a good
interpolation. Typical examples of that are the coupling
constants of the effective chiral Lagrangian of QCD at low
energies, as well as the coupling constants of the effective
chiral Lagrangian of the electroweak interactions of pseudo--scalar
particles in the Standard Model. It has been shown in some
examples~\cite{KPdeR98,KPPdeR99,PdeR00,GP00,KPdeR01} that
inserting the hadronic spectrum of large--$N_c$ QCD  as an
approximation to the real hadronic spectrum provides rather good
results. It is in this sense that large--$N_c$ QCD seems to be a
very useful phenomenological approach for understanding
non--perturbative QCD physics at low energies.

In most cases of interest, the Green's functions which govern the
low--energy constants of the chiral Lagrangian are two--point
functions with zero-momentum insertions of vector, axial--vector,
scalar and pseudo--scalar currents. The higher the power in the
chiral expansion, the higher the number of insertions.
Furthermore, they are order parameters of spontaneous chiral
symmetry breaking; i.e. they vanish, in the chiral limit where
the light quark masses are set to zero, order by order in the
perturbative vacuum of QCD. That implies that they have a
power--like fall--off in $1/Q^2$ at large $Q^2$; where
$Q^2=-q^2\ge 0$ in the euclidean region, with $q$ the
four--momentum flowing through the Green's function. When
$N_c\rightarrow \infty $, within a finite radius in the complex
$Q^2$ plane centered at the origin, these Green's functions only
have a {\it finite number of poles}.  The restriction to the
minimal number of poles required to satisfy the known {\it
short--distance} and {\it long--distance} QCD constraints results
in a simple approximation which we call the {\it minimal hadronic
approximation} (MHA) to large--$N_c$ QCD. In the cases which we
have been able to test, this minimal approximation gives already
very good phenomenological results. The MHA is in principle
improvable: considering more terms in the OPE of the two currents
in the underlying Green's function through which the
$q$--momentum flows provides extra algebraic sum rules which can
be used to fix the extra hadronic parameters. Also, knowing more
terms at small--$Q^2$ values, in the chiral expansion of the same
underlying Green's function, provides extra algebraic sum rules
of a similar nature. Unfortunately, in most cases of interest, it
is difficult to carry on these improvements in practice, and the
nature of the MHA in real QCD remains unclear. This is the reason
to resort to {\it simple models}, where the full
large--$N_c$--like spectrum is fixed {\it a priori} and one can
study approximations and their improvements explicitly. In this
paper we shall study a model defined by a spectrum of an infinity
of zero-width vector particles whose masses (squared) are equally
spaced. We consider this spectrum quite realistic. As a matter of
fact, for large masses, this \emph{is} the spectrum of QCD at
large $N_c$ in 2 dimensions (where the theory can be solved
\cite{2D}) and, in the case of our four-dimensional world, this
is also the spectrum of the daughter trajectories of Regge
Theory; which has long been suspected to be a good effective
description of QCD in the large-$N_c$ limit.

To the best of our knowledge, this type of model was
first proposed in the context of QCD in ref.~\cite{BEG72}. More
recent studies confronting phenomenology can be found in
ref.~\cite{GESH89,GP01}. Also recently Shifman \cite{SHIF00} has
emphasized the usefulness of this model for investigating the
question of {\it local duality} versus {\it global duality}.
The model has
very intriguing connections with functions which appear profusely
in analytic number theory.

The paper is organized as follows.  In section 2, we introduce the
model with equally spaced, infinitely narrow resonances.  We then
take the vector-current two-point function (more precisely, the
Adler function) as a laboratory, and calculate it in this model,
including its OPE and chiral expansions.
In section 3 we
consider a specific form of the minimal hadronic approximation.
After some discussion of its nature, we give explicit numerical
examples of how the MHA approximates the Adler function of our model.
Finally, in section 4, we compare the hadronic branching ratio
for $\tau$ decay between the model, the OPE and the MHA,
in order to see how the MHA performs with respect to local
duality and finite--energy sum rules.

\vspace*{1.0cm}
\section{\normalsize The model}

Let us consider the  vector--vector correlator defined
by
\be\lbl{2pf} i\int d^4x e^{iq\cdot x}\langle 0\vert
T\left(J^{\mu}(x),J^{\nu}(0)\right)\vert
0\rangle=(q^{\mu}q^{\nu}-g^{\mu\nu}q^2)\Pi(Q^2 \equiv -q^2)\,,
\ee
with $J^{\mu}$ the electromagnetic current of light quarks, for
example. The model for this correlator consists of
an infinite number of equally spaced narrow states as illustrated in
Fig.~1 below. With
$M_{0}$ the mass of the lowest state and $\sigma$ the spacing mass, the
spectrum is given by
 \be\lbl{spectral} \frac{1}{\pi}\Imm\Pi(t)=A\
\sigma^2\sum_{n=0}^{\infty} \delta(t-M_{0}^2-n\sigma^2)\,, \ee
with $A$ an arbitrary normalization factor that will be
determined below.

\vskip 3pc \centerline{\epsfbox{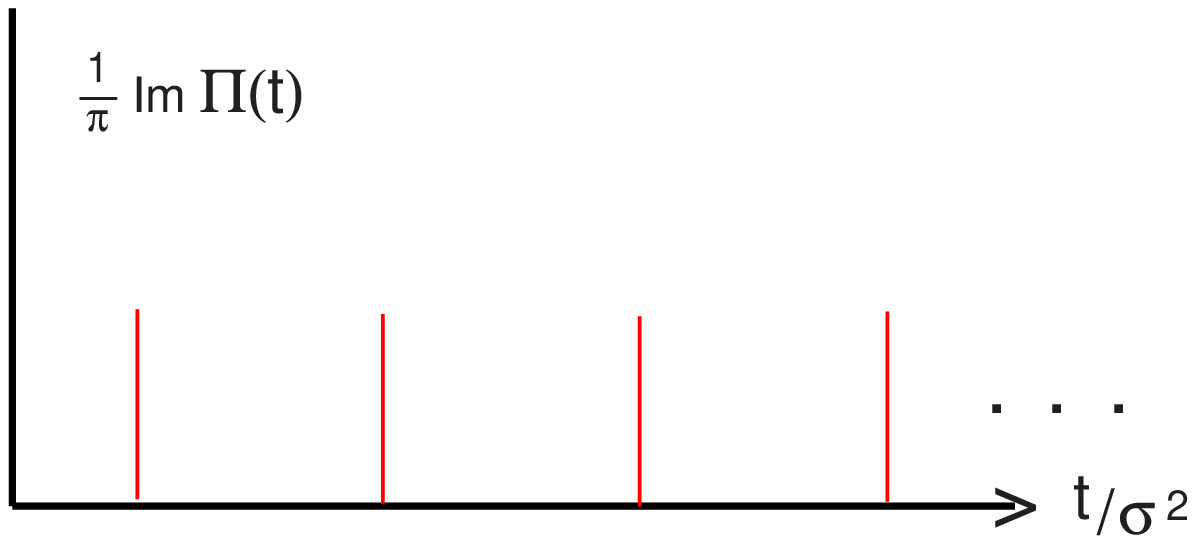}} \vskip -2pc
\centerline{{\bf Fig.~1} {\it The vector spectral function in the
toy model of large--$N_c$ QCD.}} \vskip 2pc

Of special interest for our discussion is the Adler function
defined as follows:

{\setl \bea\lbl{adler} \cA(Q^2) & = & -Q^2\frac{d}{dQ^2}
\Pi(Q^2)=\int_{0}^{\infty} dt\frac{Q^2}{(Q^2
+t)^2}\frac{1}{\pi}\Imm\Pi(t)
\\  & = &  A\ \frac{Q^2}{\sigma^2}
\sum_{n=0}^{\infty}\frac{1}{\left(\frac{Q^2}{\sigma^2}+
\frac{M_{0}^2}{\sigma^2}+n\right)^2} \equiv A\
\frac{Q^2}{\sigma^2}\zeta\left(2,\frac{Q^2+M_{0}^2}{\sigma^2}
\right)\,, \eea}

\noi where $\zeta\left(2,\frac{Q^2+M_{0}^2}{\sigma^2}\right)$ is
the so-called {\it generalized Riemann zeta function}~\cite{WW},
sometimes called the {\it Hurwitz function}\footnote{It also
coincides with $\psi'(\frac{Q^2+M_{0}^2}{\sigma^2})$, where
$\psi'(z)=\frac{d^2}{dz^2}\log \Gamma(z)$ and $\Gamma(z)$ is
Euler's Gamma  function. }. We can choose the normalization factor
$A$ so that the large--$Q^2$ behaviour of the Adler function
reproduces the parton like {\it asymptotic freedom} behaviour
predicted by perturbative QCD. This can be readily seen using the
Euler--Maclaurin formula:
\bea\lefteqn{ \frac{1}{\left(\frac{Q^2}{\sigma^2}+
\frac{M_{0}^2}{\sigma^2}\right)^2}+\frac{1}{\left(\frac{Q^2}
{\sigma^2}+ \frac{M_{0}^2}{\sigma^2}+1\right)^2}+\cdots \frac{1}
{\left(\frac{Q^2}{\sigma^2}+
\frac{M_{0}^2}{\sigma^2}+N\right)^2}=} \\ & &
\int_{0}^{N}dx\frac{1}{\left(\frac{Q^2}{\sigma^2}+
\frac{M_{0}^2}{\sigma^2}+x\right)^2}+\frac{1}{2}
\left[\frac{1}{\left(\frac{Q^2}{\sigma^2}+
\frac{M_{0}^2}{\sigma^2}\right)^2}+\frac{1}
{\left(\frac{Q^2}{\sigma^2}+ \frac{M_{0}^2}{\sigma^2}+N\right)^2}
\right]+\cdots\,, \eea and the integral result \be
\int_{0}^{N}dx\frac{1}{\left(\frac{Q^2}{\sigma^2}+
\frac{M_{0}^2}{\sigma^2}+x\right)^2}=\frac{-1}{\frac{Q^2}{\sigma^2}+
\frac{M_{0}^2}{\sigma^2}+N}-\frac{-1}{\frac{Q^2}{\sigma^2}+
\frac{M_{0}^2}{\sigma^2}}\,. \ee Letting $N\ra\infty$ we get the
result \be \lim_{Q^2\ra\infty}\cA(Q^2)\Ra A\,, \ee which, for
$A=\frac{N_c}{16\pi^2}\frac{4}{3}$, reproduces the asymptotic
parton limit result. For simplicity, we shall take $A=1$ from
here onwards.

There is an integral representation of the Hurwitz function which
will be very useful for our purposes:
\be\lbl{repzeta} \zeta(s,z)=\frac{1}{\Gamma(s)}\int_{0}^\infty dt
\,e^{-zt}\,\frac{t^{s-1}}{1-e^{-t}}\,\qquad (\Ree\ s>
1\quad\annd\quad \Ree\ z>0)\,. \ee
\vspace*{0.5cm}
\subsection{\normalsize\sc  Operator Product Expansion of the model}

\noi In QCD, the behaviour of the Adler function in the physical
vacuum, for large euclidean $Q^2$ values, can be obtained from
the operator product expansion (OPE) of the two currents in
Eq.~\rf{2pf} and it results in an expansion in $1/Q^2$ powers
\`{a} la Shifman, Vainshtein, Zakharov~\cite{SVZ79}. In the model
this expansion follows readily from the integral representation
\rf{repzeta} applied to
$\zeta\left(2,\frac{Q^2+M_{0}^2}{\sigma^2}\right)$.

Using the fact that \be
e^{xz}\frac{z}{e^z
-1}=\sum_{n=0}^{\infty}B_{n}(x)\frac{1}{n!}z^n\,, \ee is the
generating function of the Bernoulli polynomials,\footnote{For
the sake of completeness we list a few useful properties of the
Bernoulli polynomials in an Appendix.} we have \be\lbl{bergim}
e^{-\frac{M_{0}^2}{\sigma^2}t}\frac{t}{1-e^{-t}}=
e^{\left(1-\frac{M_{0}^2}{\sigma^2}\right)t}\frac{t}{e^t -1}=
\sum_{n=0}^{\infty}B_{n}\left(1-\frac{M_{0}^2}{\sigma^2}\right)
\frac{1}{n!}t^n\,, \ee and using the symmetry property in
Eq.~\rf{half}, plus the Laplace transform integral representation \be
\frac{1}{s^{n+1}}=\frac{1}{\Gamma(n+1)}\int_{0}^{\infty}dt\,
e^{-st}t^n\,, \ee we get the required OPE--like expansion

\be\lbl{OPEex} \cA^{(\infty)}(Q^2)=\sum_{n=0}^{\infty}(-1)^n B_{n}
(\hat{M}_{0}^2)
\left(\frac{\sigma^2}{Q^2}\right)^n\,,\quad\with\quad
\hat{M}_{0}^2\equiv\frac{M_{0}^2}{\sigma^2}\,. \ee
The explicit
expansion for the first few terms is then
{\setl
\bea\lbl{eq:19}
\lefteqn{\mathcal{A}^{(\infty) }({Q^2})
  \rightarrow 1+ \left(\frac{1}{2}-\ \hat{M_0}^2
    \right)\left( \frac{\sigma^2}{Q^2} \right) +\left(\frac{1}{6}-
\hat{M_0}^2+{\hat{M_0}^4}\right) \left(
      \frac{\sigma^2}{{Q^2}} \right)^2} \\
& &  \!\!\!+
  \left(-\frac{\hat{M_0}^2}{2}+\frac{3{\hat{M_0}^4}}{2}-{\hat{M_0}^6}\right)
\left(\frac{\sigma^2}{{Q^2}}
  \right)^3 +\left(-\frac
    {1}{30}+{\hat{M_0}^4}-2{\hat{M_0}^6}+{\hat{M_0}^8}\right)
\left( \frac{\sigma^2}{{Q^2}} \right)^4 \!+
\cO\left(\frac{\sigma^2}{Q^2}\right)^5\!\!\!\,. \nonumber
\eea}

In QCD, and in the chiral limit where the light quark masses are
set to zero, there is no $1/Q^2$ term in the OPE, because there
is no local operator of dimension $d=2$. We can then make the
model look like QCD more closely by imposing this property in
Eq.~\rf{eq:19} as well, which requires that \be\lbl{D2R}
M_{0}^2=\frac{1}{2}\sigma^2\,. \ee The model has then only one
mass scale $\sigma^2$; furthermore, because of the property of
the Bernoulli polynomial that \be B_{2n+1}(1/2)=0\,,\qquad \foor
\qquad n\ge 0\,, \ee the equivalent of the OPE of the Adler
function in the model, has only {\it even} $1/Q^2$ powers: \be
\cA^{(\infty)}(Q^2)\ra 1-\frac{1}{12}\left( \frac{\sigma^2}{Q^2}
\right)^2+\frac{7}{240}\left( \frac{\sigma^2}{Q^2} \right)^4-
\frac{31}{1344}\left( \frac{\sigma^2}{Q^2} \right)^6+\cdots\,, \ee
and the vacuum expectation values appearing in the OPE are then
given by the values of the Bernoulli polynomials at $x=1/2$. For
the rest of the paper we shall keep $M_{0}^2=\sigma^2/2$ and
comment on the relevance of this constraint whenever necessary.

 It is easy to see that the OPE series of the model in question
is a divergent series. From Eq.~\rf{bertrigoeven} it follows that
\be B_{2N}(1/2)=(-1)^{N+1}
2(2N)!\frac{1}{(2\pi)^{2N}}\sum_{n=1}^{\infty}\frac{(-1)^n}{n^{2N}}\,,
\ee
and the  sum of the series on the right--hand side is known:
\be
\sum_{n=1}^{\infty}\frac{(-1)^n}{n^{2N}}=\frac{(1-2^{2N-1})\pi^{2N}}
{(2N)!}\vert B_{2N}\vert\,, \ee which results in \be
B_{2N}(1/2)=(-1)^{N}\left(1-\frac{1}{2^{2N-1}} \right) \vert
B_{2N}\vert\,,\ \ \ N>0\,. \ee From this, it is easy to see that
the series is divergent because \be \lim_{N\ra\infty}\frac{\vert
B_{2N}(1/2)\vert \left(\frac{\sigma^2}{Q^2} \right)^{2N}}{\vert
B_{2N-2}(1/2)\vert \left(\frac{\sigma^2}{Q^2} \right)^{2N-2}}  =
\lim_{N\ra\infty}\frac{\vert B_{2N}\vert}{\vert B_{2N-2}\vert}
\left(\frac{\sigma^2}{Q^2} \right)^{2}   \sim
\left(\frac{N}{\pi}\frac{\sigma^2}{Q^2} \right)^2\,, \ee where we
have used the property that asymptotically, \be
\lim_{N\ra\infty}\vert B_{2N}\vert\sim
4\sqrt{\pi}\frac{N^{2N+1/2}}{(e\pi)^{2N}}\,. \ee At orders
\be\lbl{asope} 2N\gtrsim 2\pi\frac{Q^2}{\sigma^2}\,, \ee the OPE
ceases to be an improving expansion.

The OPE series  is divergent, but summable. In fact, the Borel
sum of the OPE series is precisely the Adler function defined by
the generalized Riemann zeta function. To see this, recall that
the Borel sum of the series in Eq.~\rf{OPEex} is defined as
follows \be\lbl{borel} \sum_{n=0}^{\infty}(-1)^n B_{n} (1/2)
\left(\frac{\sigma^2}{Q^2}\right)^n\ra\int_{0}^{\infty}dt e^{-t}
\sum_{n=0}^{\infty}(-1)^n \frac{B_{n} (1/2)}{n!}
\left(\frac{\sigma^2}{Q^2}t\right)^n\,. \ee If now, we use again
the symmetry property in Eq.~\rf{half} and the second
of Eq.~\rf{bergim}; plus the change of variables
$\tilde{t}=t\frac{\sigma^2}{Q^2}$, one obtains the desired result.

\vspace*{0.5cm}

\subsection{\normalsize\sc  Chiral Perturbation Theory expansion of the
model}

\noi We can also find, in this model, the equivalent of the chiral
perturbation theory ($\chi$PT)
expansion  of the  Adler function; i.e., the Taylor
expansion at small $Q^2$ values. For that, it is sufficient to
expand the term $e^{-\frac{Q^2}{\sigma^2}t}$ in the integral
representation of Eq.~\rf{repzeta} with the result
\be \cA^{(0)}(Q^2)=\sum_{n=0}^{\infty}(-1)^n\left(
\frac{Q^2}{\sigma^2}
\right)^{n+1}(n+1)\,\,\,\zeta\left(n+2,\frac{M_{0}^2}{\sigma^2}
\right)\,. \ee With $M_{0}^2=\frac{1}{2}\sigma^2$, and using the
property that \be \zeta(s,1/2)=(2^s -1)\, \zeta(s)\,, \qquad
\foor\qquad s>1\,, \ee
 the
series becomes
\be\lbl{chiral}
\cA^{(0)}(Q^2)=\sum_{n=0}^{\infty}(-1)^{n}\,\,(n+1)\left[2^{(n+2)} -1
\right]
\zeta(n+2)\left(
\frac{Q^2}{\sigma^2}
\right)^{n+1}\,,
\ee
i.e., the low--energy constants of the Adler function are
proportional to Riemann's zeta functions of increasing
argument.
The explicit chiral expansion for the first few terms gives
\be
\cA^{(0)}(Q^2)\ra \frac{\pi^2}{2}\left(
\frac{Q^2}{\sigma^2}
\right)-14\,\zeta(3)\left(\frac{Q^2}{\sigma^2}\right)^2+\frac{\pi^4}{2}
\left(\frac{Q^2}{\sigma^2}\right)^3-124\,\zeta(5)\left(\frac{Q^2}
{\sigma^2}\right)^4+\cO\left(
\frac{Q^2}{\sigma^2}
\right)^5\,.
\ee
Numerically,
\be
\cA^{(0)}(Q^2)\ra
4.9348\left(\frac{Q^2}{\sigma^2}\right)-16.8288
\left(\frac{Q^2}{\sigma^2}\right)^2+48.7045
\left(\frac{Q^2}{\sigma^2}\right)^3-128.579
\left(\frac{Q^2}{\sigma^2}\right)^4+\cdots\,.
\ee
The chiral expansion is convergent within a radius
$\frac{Q^2}{\sigma^2}\le \frac{1}{2}$.

It is indeed amusing that a model of a well defined limit of a rather
complicated quantum field theory like QCD, has such direct
connections with the basic functions of analytic number theory and
transcendental numbers.

\vspace*{1.0cm}
\section{\normalsize The Minimal Hadronic Approximation.}

\noi We want to study how good is the approximation which
consists in replacing the exact spectral function in
Eq.~\rf{spectral} by a finite set of low states, say $m+1$, with
arbitrary masses $\tilde{M}_{n}^2$, and lumping the rest
into the equivalent of a perturbative QCD continuum
starting at a mass squared value $s_{0}$:
\be\lbl{mhasf}
\frac{1}{\pi}\Imm\Pi(t)\Big\vert_{\mbox{\tiny\rm MHA}}^{(m+1)}=
\sigma^2\sum_{n=0}^{m}
\delta(t-\tilde{M}_{n}^2)+\theta(t-s_{0})\,. \ee The approximated
spectrum is illustrated in Fig.~2. We call this the {\it minimal
hadronic approximation} (MHA), by analogy to the approximations
which we have been making in real large--$N_c$ QCD.\footnote {See
ref.~\cite{deR01} for a recent review, and refs.~\cite{KPdeR98}
to \cite{KPdeR01} for details.}

\vskip 3pc \centerline{\epsfbox{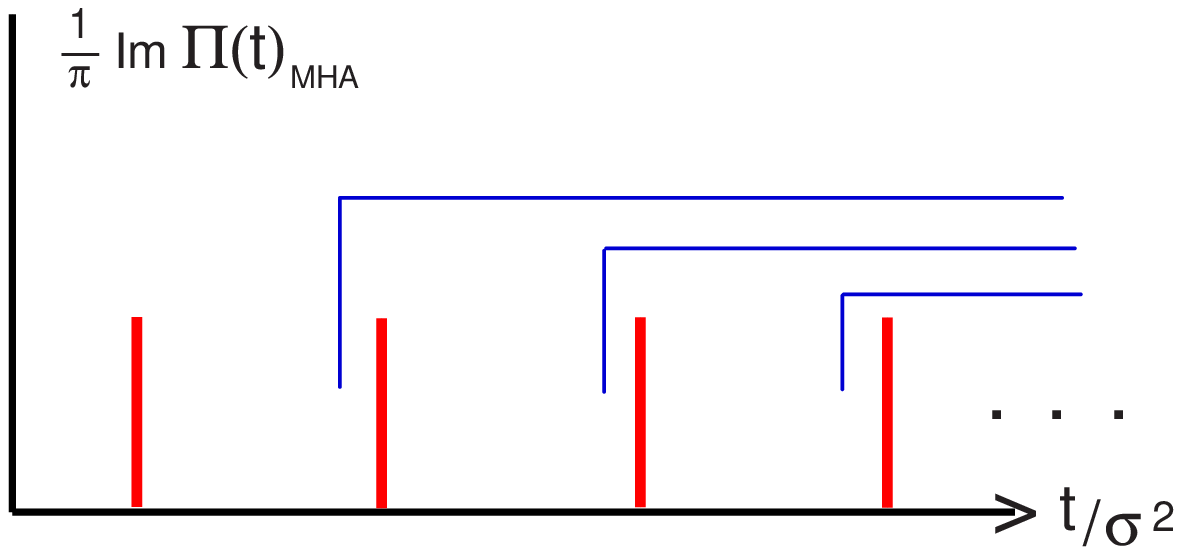}} \vskip -2pc {\bf
Fig.~2} {\it The Minimal Hadronic Approximation (MHA) to the
vector spectral function in the toy model of large--$N_c$ QCD.
The successive approximations correspond to one state plus
continuum (the first blue band), two states plus continuum (the
second blue band), three states plus continuum, etc.} \vskip 2pc
The MHA spectrum leads to the following form for the Adler
function: \be\lbl{MHA} \cA(Q^2)\Big\vert_{\mbox{\tiny\rm
MHA}}^{(m+1)}=\frac{Q^2}{\sigma^2}
\sum_{n=0}^{m}\frac{1}{\left(\frac{Q^2}{\sigma^2}+
\frac{\tilde{M}_{n}^2}{\sigma^2}\right)^2}+\frac{Q^2}
{Q^2+s_{0}}\,, \ee where the second term is the one induced by
the perturbative continuum. The OPE--like expansion of the MHA
approximation to the toy model can be easily obtained from the
expression above, with the result \be\lbl{opemha}
\cA^{(\infty)}(Q^2)\vert_{\mbox{\tiny\rm
MHA}}=1+\sum_{N=1}^{\infty}
(-1)^{N}\left[\left(\frac{s_{0}}{\sigma^2}\right)^{N}-N\sum_{n=0}^{m}
\left(\frac{\tilde{M}_{n}^2}{\sigma^2} \right)^{N-1}
\right]\left(\frac{\sigma^2}{Q^2} \right)^{N}\,. \ee

We fix the onset of the continuum $s_{0}$ to match the leading
$1/Q^2$ term of the OPE for the Adler function, which vanishes.
This results in the condition\footnote{If $\hat M_0^2\neq 1/2$, Eq.
\rf{cont} reads $s_0=\sigma^2 (m+1/2+\hat M_0^2)$. Notice that
$s_0$ should always be larger than the mass of the last resonance
kept explicitly in the spectrum.} \be\lbl{cont}
s_{0}=(m+1)\sigma^2\,. \ee

 The masses $\tilde{M}_{N}^2$ of the explicit narrow states are
then fixed so as to satisfy more and more constraints provided by
the OPE and the $\chi$PT expansions of the underlying theory.
This results in a system of equations: \be\lbl{systemope}
(m+1)^N-N\sum_{n=0}^{m} \left(\frac{\tilde{M}_{n}^2}{\sigma^2}
\right)^{N-1}=B_{N}(1/2) \,,\qquad\foor\qquad N=2,3,\dots, \ee
from the OPE constraints; and another system of equations

\bea\lbl{systelch} \frac{1}{N(m+1)^N}+\sum_{n=0}^{m}\left(
\frac{\sigma^2}{\tilde{M}_{n}^2}
\right)^{N+1}=&&\zeta(N+1,\frac{1}{2})\\ \nn
=&&\sum_{n=0}^\infty\frac{1}{(n+\frac{1}{2})^{N+1}}\ ,
\quad\mathrm{for}\quad  N=1,\dots, \eea
from the $\chi$PT constraints. The idea is to solve these
equations for successive values of the number $m$ of explicit
narrow states. Looking at Eq. \rf{systelch} one sees that the
first term on the left--hand side vanishes in the $m\to\infty$
limit.  It is clear that, in this limit, the system of equations
\rf{systelch} has as a solution $\tilde M_n^2=\sigma^2(n+1/2)$,
i.e. the true spectrum.

We also find (numerically, including up to nine resonances) that
when one substitutes the masses found from only the $\chi$PT
constraints into the left--hand side of eq.~\rf{systemope},
the resulting numbers converge to the values given by the
right--hand side (i.e. the $B_N(1/2)$).  This implies that
in principle, the MHA will converge to the exact result for
{\em all} values of $Q^2$, even if one uses {\em only} small $Q^2$
information.  However, the convergence is rather
slow for large $Q^2$.
Therefore, in practice, one would expect better convergence
if one also uses constraints coming from the OPE in determining
the parameters $\tilde{M_n^2}$ of the MHA.

It is less clear what the nature is of the system of equations
following from using only OPE constraints since, unlike the
$\chi$PT  expansion, the OPE is an asymptotic expansion (see
section (2.1)). For a given number of OPE constraints, $2N$,  Eq.
\rf{asope} gives a value of $Q^2$ below which the MHA
obtained by solving the set of equations
\rf{systemope} may be wrong.\footnote{Although not
\emph{necessarily} so. The Stieltjes function
$\int_{0}^{\infty}dt\frac{e^{-t}}{1+zt}$, which has
$\sum_{n=0}^{\infty} (-1)^n n! z^n$ as the asymptotic series for
$z\rightarrow 0$, can actually be approximated in a convergent
sequence of steps with this method \cite{Bender}.}  Conversely,
given a value of $Q^2$ above which one wishes to obtain a good
approximation to the exact model, eq.~\rf{asope} gives an
estimate of the maximum useful order of  the OPE.  However, as
already alluded to above, we find it quite plausible to expect
that using a few constraints coming from the OPE in addition to
those from $\chi$PT should speed up the convergence in the region
of \emph{large} $Q^2$, relative to a MHA solely based on $\chi$PT
constraints. In practice, quantities which typically one encounters in
QCD correspond to integrals of the Adler
function weighted by a known function, and inspection of the
weight function should dictate the optimal choice of short--distance
constraints versus long--distance constraints. For illustration
purposes, in the  examples which follow below,  we shall always combine
both short-- and long--distance constraints.  An important comment,
however, is that when one uses only chiral constraints, the masses
$\tilde M_n$ appear to converge monotonously to the exact values, while
if one also uses OPE constraints, this appears not to be true.  Even so,
the approximate Adler function, in all the numerical experiments we have
examined, always gets better for all values of $Q^2$ if one includes more
resonances in a systematic way.

We shall now construct explicitly the leading and next-to-leading
Minimal Hadronic Approximation to the model.

\vspace*{0.5cm}
\subsection{\normalsize\sc One State plus Continuum}

\noi
The onset of the continuum is now
\be
s_{0}=\sigma^2\,,
\ee
and the first OPE constraint becomes
\be
1-2\frac{\tilde{M}_{0}^2}{\sigma^2}  =
-\frac{1}{12}\,.
\ee
The numerical value of the solution is
\be
\tilde{M}_{0}^2=0.541667\sigma^2\,,
\ee
not quite the exact value $M_{0}^2=\frac{1}{2}\sigma^2$ of the
{\it real spectrum} but not bad at all! In fact, as shown in Fig.~3,
the simple MHA of one state plus continuum, already reproduces rather
well the shape of the {\it exact} Adler function.  For comparison,
if $\tilde M_0^2$ is fixed using the first $\chi$PT constraint,
one finds $\tilde M_0^2=0.504125\sigma^2$.  While this is much closer
to the exact value $M_0^2=0.5\sigma^2$, the approximate Adler
function turns out to be a worse approximation for
$Q^2\gtrsim1.8\sigma^2$.

\vskip 3pc \centerline{\epsfbox{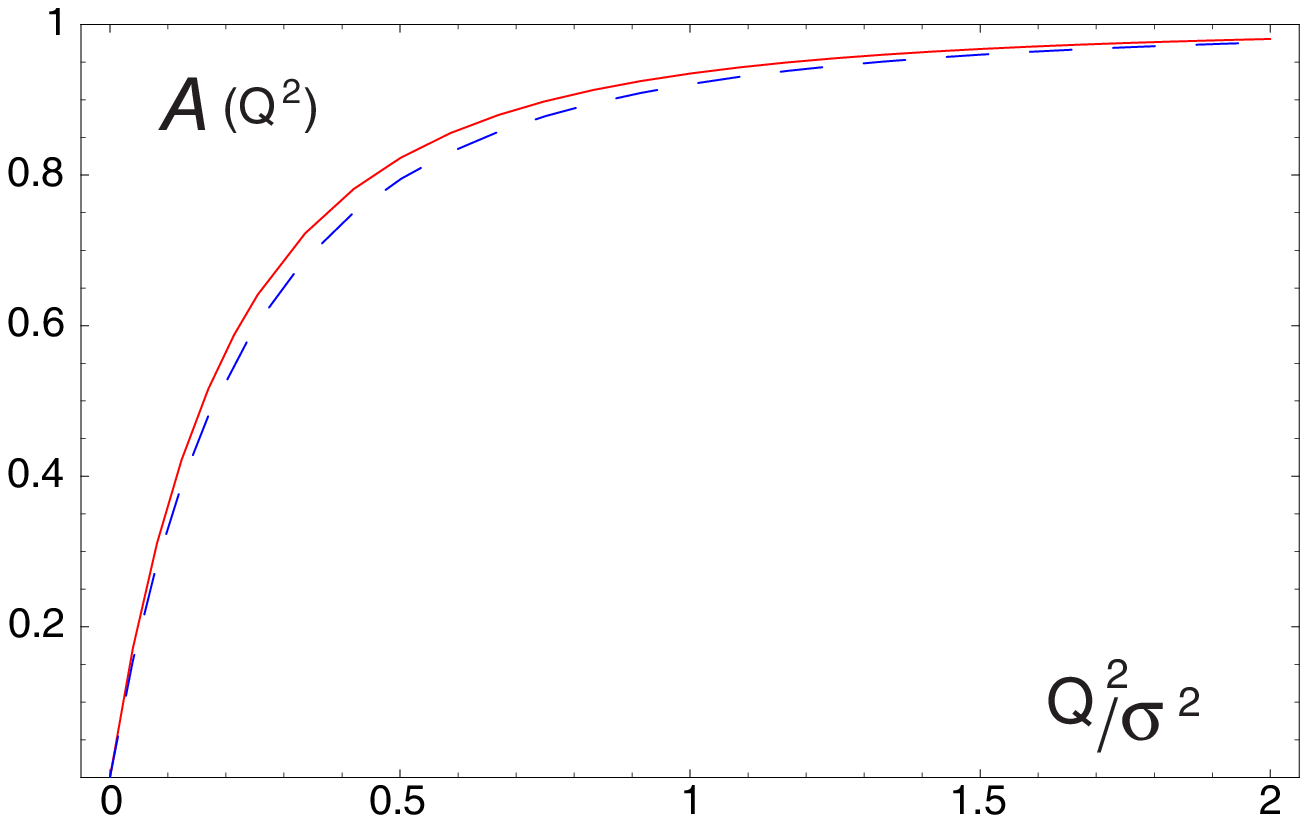}} \vskip 1pc {\bf
Fig.~3} {\it The exact Adler function of the toy model (the
continuous red curve), and the Adler function predicted by the
MHA with one state plus continuum (the dashed blue curve), are
plotted versus $Q^2/\sigma^2$. } \vskip 2pc

\vspace*{0.5cm}
\subsection{\normalsize\sc Two States plus Continuum}

\noi
We can now tune the masses $\tilde{M}_{0}$ and $\tilde{M}_{1}$ to
reproduce the leading $\frac{1}{Q^4}$ and  next--to--leading
$\frac{1}{Q^6}$ terms of the underlying OPE. The constraining equations
are now

{\setl
\bea\lbl{2states}
4-2\frac{\tilde{M}_{0}^2}{\sigma^2}-2\frac{\tilde{M}_{1}^2}{\sigma^2}
& = & \frac{-1}{12}\,, \\
-8+3\frac{\tilde{M}_{0}^4}{\sigma^4}+3\frac{\tilde{M}_{1}^4}{\sigma^4}
& = & 0\,.
\eea}

\noi
They lead to the solutions:
\be\lbl{masses220}
\tilde{M}_{0}^2=0.481174\sigma^2\qquad\annd\qquad
\tilde{M}_{1}^2=1.56049\sigma^2\,,
\ee
instead of the physical values $M_{0}^2=\frac{1}{2}\sigma^2$ and
$M_{1}^2=\frac{3}{2}\sigma^2$. Notice that the mass of the ground
state in the two state approximation gets closer to the physical mass.
The corresponding Adler function also approaches better the {\it exact}
Adler function. In order to quantify this improvement we plot in
Fig.~4 the difference of functions
\be\lbl{dif12}
\Delta^{(1)}\equiv\cA(Q^2)-\cA(Q^2)\Big\vert_{\mbox{\tiny\rm
MHA}}^{(1)}\quad\annd\quad
\Delta^{(2)}\equiv\cA(Q^2)-\cA(Q^2)\Big\vert_{\mbox{\tiny\rm
MHA}}^{(2)}\,,
\ee
versus $Q^2$.

\vskip 3pc \centerline{\epsfbox{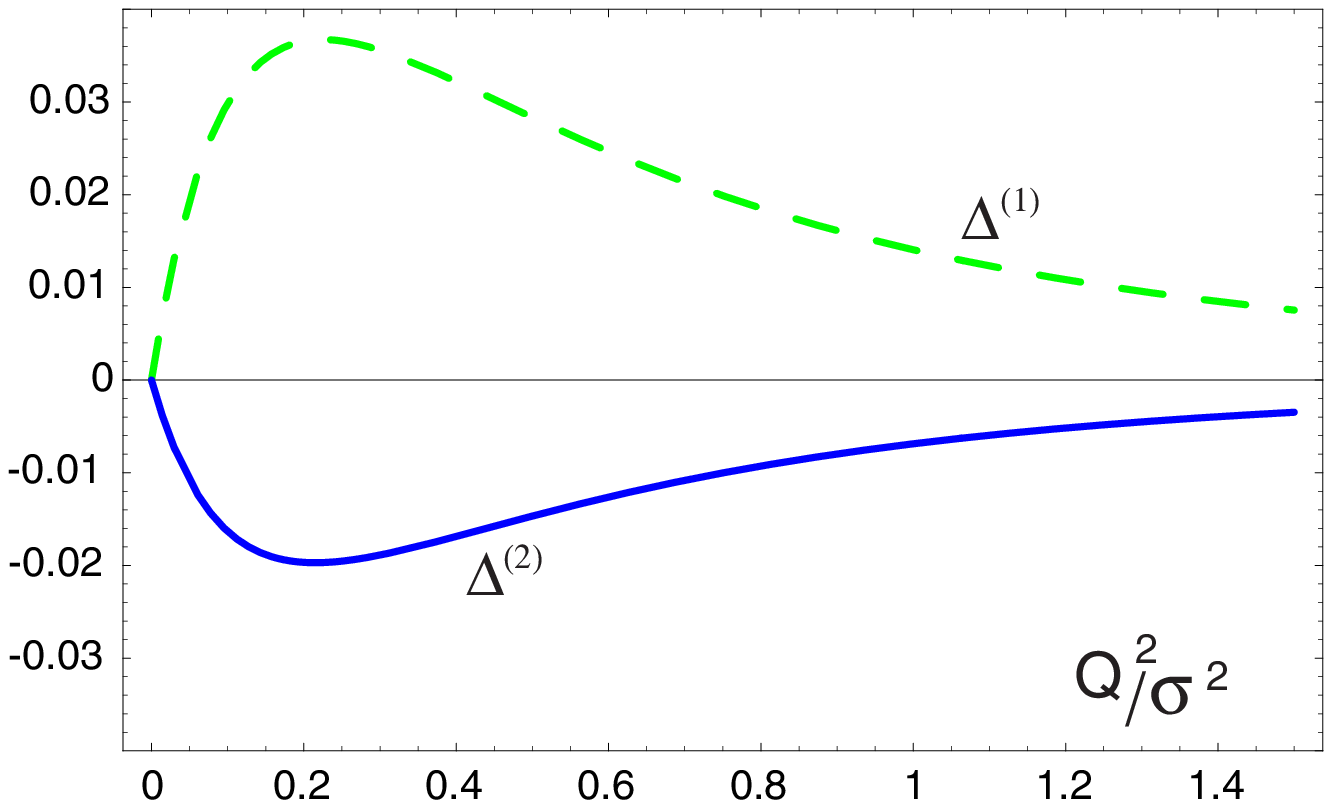}} \vskip 1pc {\bf
Fig.~4} {\it The difference of Adler functions defined in
Eq.~\rf{dif12} corresponding to the MHA with one state plus
continuum (the dashed green curve) and with two states plus
continuum (the continuous blue curve) are plotted versus
$Q^2/\sigma^2$. Notice the vertical scale in the plot. } \vskip
2pc

Another alternative is to use as constraints the leading
$\frac{1}{Q^4}$ behaviour of the OPE and the leading $Q^2$
behaviour of the chiral expansion.  The constraining equations
are then

{\setl
\bea\lbl{2states11}
4-2\frac{\tilde{M}_{0}^2}{\sigma^2}-2\frac{\tilde{M}_{1}^2}{\sigma^2}
& = & \frac{-1}{12}\,, \\ \lbl{1stld}
\frac{1}{2}+\frac{\sigma^4}{\tilde{M}_{0}^4}+
\frac{\sigma^4}{\tilde{M}_{1}^4}
& = & \frac{\pi^2}{2}\,;
\eea}

\noi and the corresponding numerical solutions for the masses are
then: \be\lbl{masses211}
\tilde{M}_{0}^2=0.499093\sigma^2\qquad\annd\qquad
\tilde{M}_{1}^2=1.54257\sigma^2\,. \ee They approach, even
better, the {\it exact} masses than those previously obtained in
Eq.~\rf{masses220}. In fact, combining the leading
short--distance constraint with the leading long--distance
constraint turns out to produce an approximated Adler function
much closer to the exact one. We illustrate this in Fig.~5 where
the dashed blue curve is the same as the continuous blue curve in
Fig.~4 (though plotted on a different scale), while the continuous
red curve is the one obtained solving the constraints of
Eq.~\rf{2states11}. The improvement is rather remarkable. We
would like to emphasize that in the applications of the MHA  to
large--$N_c$ QCD that we have made so far in refs.~\cite{KPdeR98}
to \cite{KPdeR01}, we have used equivalent constraints to the
ones discussed in this example.

In Fig.~5 we use the following notation: $\Delta_{ij}^{(m)}$
indicates the difference between the exact Adler function
and the one obtained using the MHA with $m$ states plus
continuum, with the masses obtained by solving a system of
equations which consists of $i$ short--distance constraints and
$j$ long distance constraints; i.e., \be\lbl{notation}
\Delta_{ij}^{(m)}\equiv\cA(Q^2)-\cA(Q^2)\Big\vert_{\mbox{\tiny\rm
MHA}}^{(m)}\left\{\begin{array}{r}\with~i~{\mbox{\rm
SD--constraints}}\\ \annd~j~{\mbox{\rm LD--constraints}}
\end{array}\right\}\,,\quad i+j=m\,. \ee

\vskip 3pc \centerline{\epsfbox{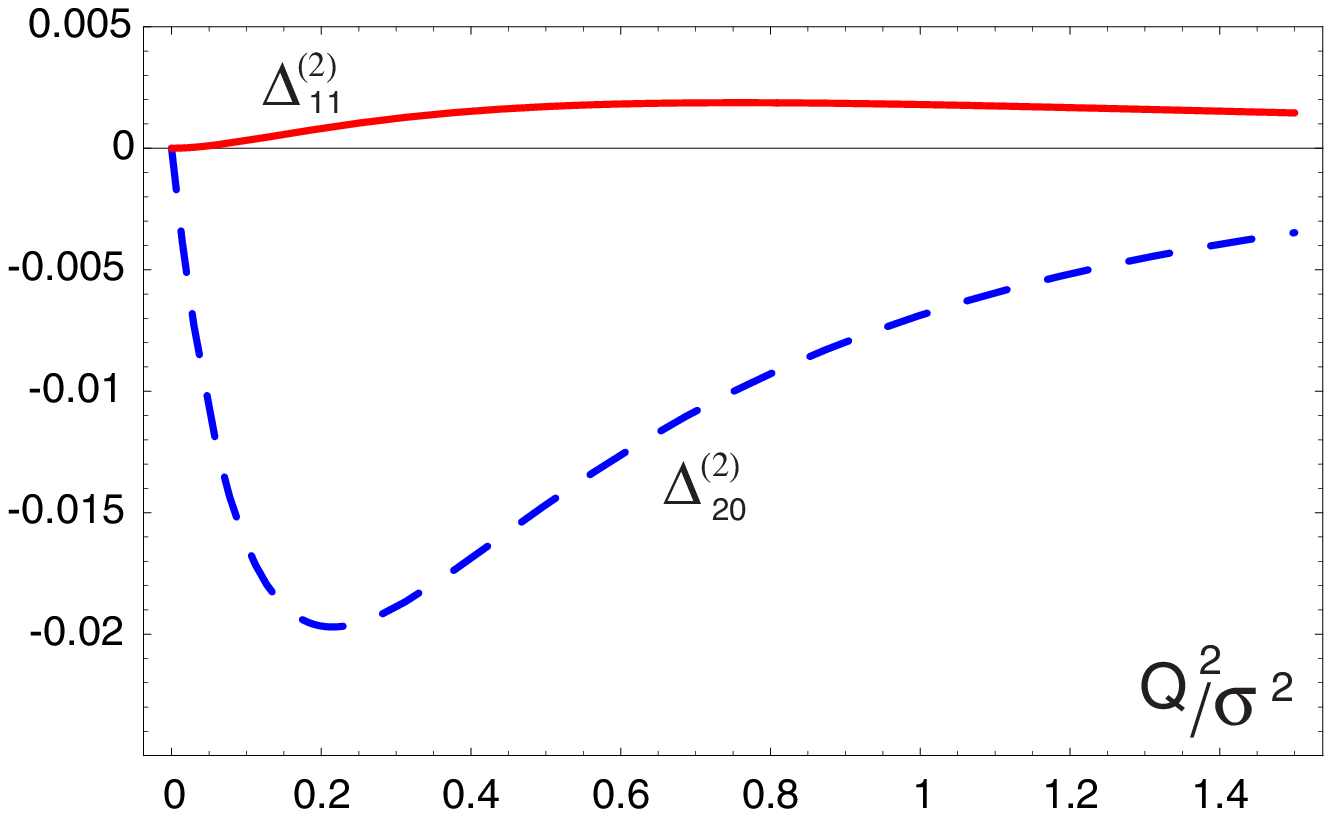}} \vskip 1pc {\bf
Fig.~5} {\it The difference of Adler functions corresponding to
the MHA  with two states plus continuum. The dashed blue curve is
the same as the one in Fig.~4, though on a different scale, while
the continuous red curve is the one obtained using as input
masses the solutions in Eq.~\rf{masses211}.} \vskip 2pc

We have constructed the MHA with up to 11 resonances. With a
similar number of short-distance  and long-distance constraints,
the maximum difference between MHA and the exact Adler function
is always decreasing by half an order of magnitude (in average) every
time a new resonance is explicitly included in the MHA. This
maximum difference occurs for $Q^2$ at roughly 1/3 of the number
of resonances included. For instance for 11 resonances this
maximum difference is of the order of $10^{-8}$ and happens at
$Q^2/\sigma^2 \simeq 4$.

\vspace*{1.0cm}
\section{\normalsize Finite--Energy Sum Rules and Hadronic Duality.}

In every day life QCD, the paradigm of a finite--energy sum rule
(FESR) is the one provided by the hadronic decay of the $\tau$,
which relates the branching ratio \be
\cR_{\tau}=\frac{\Gamma(\tau^{-}\ra \nu_{\tau}\  {\mbox{\rm
Hadrons}})} {\Gamma(\tau^{-}\ra \nu_{\tau}\
e^{-}\bar{\nu}_{e})}\,, \ee to the vector and axial--vector
hadronic spectral functions. In the chiral limit, this relation
is simply given by the integral\footnote{Here we only
consider the vector, and not the axial--vector, part.}
\be\lbl{Rtau}
\cR_{\tau}(s)=\int_{0}^{s}\frac{dt}{s}\left(1-\frac{t}{s}
 \right)^2 \left(1+2\frac{t}{s}\right)\frac{1}{\pi}\Imm\Pi(t)\,,
\ee for $s=m_{\tau}^2$. Sum rules of this type have been
extensively studied in the literature\footnote{See e.g.
refs.~\cite{BNP92,LeDP92}.} and proved very useful for extracting
the value of the QCD running coupling constant. The toy  model
offers an interesting theoretical laboratory to check these sum
rules. Some observations in this respect have already been made by
Shifman~\cite{SHIF00}. For the sake of clarity, we reproduce part
of his analysis in our discussion.

Equation \rf{Rtau} is a combination of moment sum rules of the
type \be \cM^{(p)}(s)=\int_{0}^{s}\frac{dt}{s}\left(\frac{t}{s}
\right)^{p} \frac{1}{\pi}\Imm\Pi(t)\,. \ee In the toy model, with
the input spectral function in Eq.~\rf{spectral}, these moments
can be explicitly calculated with the result

\be \cM^{(p)}(s)=\sigma^2
\sum_{n=0}^{\frac{s}{\sigma^2}-\frac{1}{2}-x}
\frac{(n+\frac{1}{2})^p\sigma^{2p}}{s^{p+1}}\,, \ee where $0\leq
x< 1$ is the fractional part of $\frac{s}{\sigma^2}-\frac{1}{2}$;
in other words, \be\lbl{x} s=(m+x+\frac{1}{2})\sigma^2\ ,\quad
\mathrm{for} \quad m=0,1,2,\ldots . \ee The exact $\cR_{\tau}(s)$
in the toy model of large--$N_c$ is then

{\setl \bea\lbl{Rexact} \cR_{\tau}(s) & = & \frac{1}{2}-
\frac{1}{2} \left(\frac{\sigma^2}{s}\right)^3 \ x (1-x) (1-2x)
  +  \frac{1}{2}\left(\frac{\sigma^2}{s}\right)^4 \left[ x^2 (1-x)^2
-\frac{1}{16}\right]\,. \eea}

\noi This expression could be viewed as the {\it exact}
$\cR_{\tau}(s)$ calculated from the {\it data} in an imaginary
world in which the $\tau$ mass may be dialed to be as large as one
wishes. This exact  $\cR_{\tau}(s)$ is plotted in Fig.~6 versus
the ratio $s/\sigma^2$ (the red continuous curve ) and as it can
be seen, it shows a damped oscillatory behaviour induced by the
convolution in the integrand of $\cR_{\tau}(s)$ in Eq. \rf{Rtau}
of a decreasing polynomial in $s$ and an infinite set of
equally-spaced  ``kicks" (i.e. the narrow resonances). Should the
number of resonances stop at a certain finite number $m$, beyond which
the continuum takes over, the oscillations for $\frac{s}{\sigma^2}> m +
\frac{1}{2}$ would disappear and  turn into an ordinary power fall-off
towards the value of the parton model, i.e. 1/2.

It is interesting to compare the exact $\cR_{\tau}(s)$ function in
Eq.~\rf{Rexact} with the one predicted by the OPE  of the
same model. Emulating what is being done in the real-world
analysis of $\tau$ decay, this comparison can be obtained from an
integral representation in the complex $z$--plane
(where $z=Q^2/s$),
{\setl \bea \cR_{\tau}(s) & = & \frac{-1}{2\pi i}\oint_{\mid
z\mid=1} dz\left(1+z
 \right)^2 \left(1-2z\right)\Pi(zs) \\ \lbl{adlerex}
 & = & \frac{1}{2\pi i}\oint_{\mid z\mid=1} \frac{dz}{z}\left(
\frac{1}{2}-z+z^3+\frac{1}{2}z^4 \right)\cA(zs)\,, \eea}

\noi by inserting the OPE for the Adler function in
Eq.~\rf{eq:19}. Doing this one notices that, besides the partonic
contribution, only the powers proportional to $1/Q^2$, $1/Q^6$
and $1/Q^8$ in the OPE of the Adler function contribute to the
integral, with the result

{\setl \bea\lbl{Rope} \cR_{\tau}^{{\mbox{\rm\tiny OPE}}}(s) & = &
\frac{1}{2}-\frac{7}{480} \left(\frac{\sigma^2}{s}\right)^4 \eea}

The curve corresponding to this $\cR_{\tau}^\mathrm{OPE}(s)$
function is also plotted in Fig.~6 (the blue dashed curve). It
interpolates rather well the {\it exact} curve but it fails to
reproduce its oscillations \cite{SHIF00}. This is another
manifestation of the violation of local duality (i.e. point--wise;
as opposed to global duality, i.e. on average) between the
quark--gluon picture represented by the OPE and the hadron picture
represented by the full spectrum--based solution.

\vskip 2pc \centerline{\epsfbox{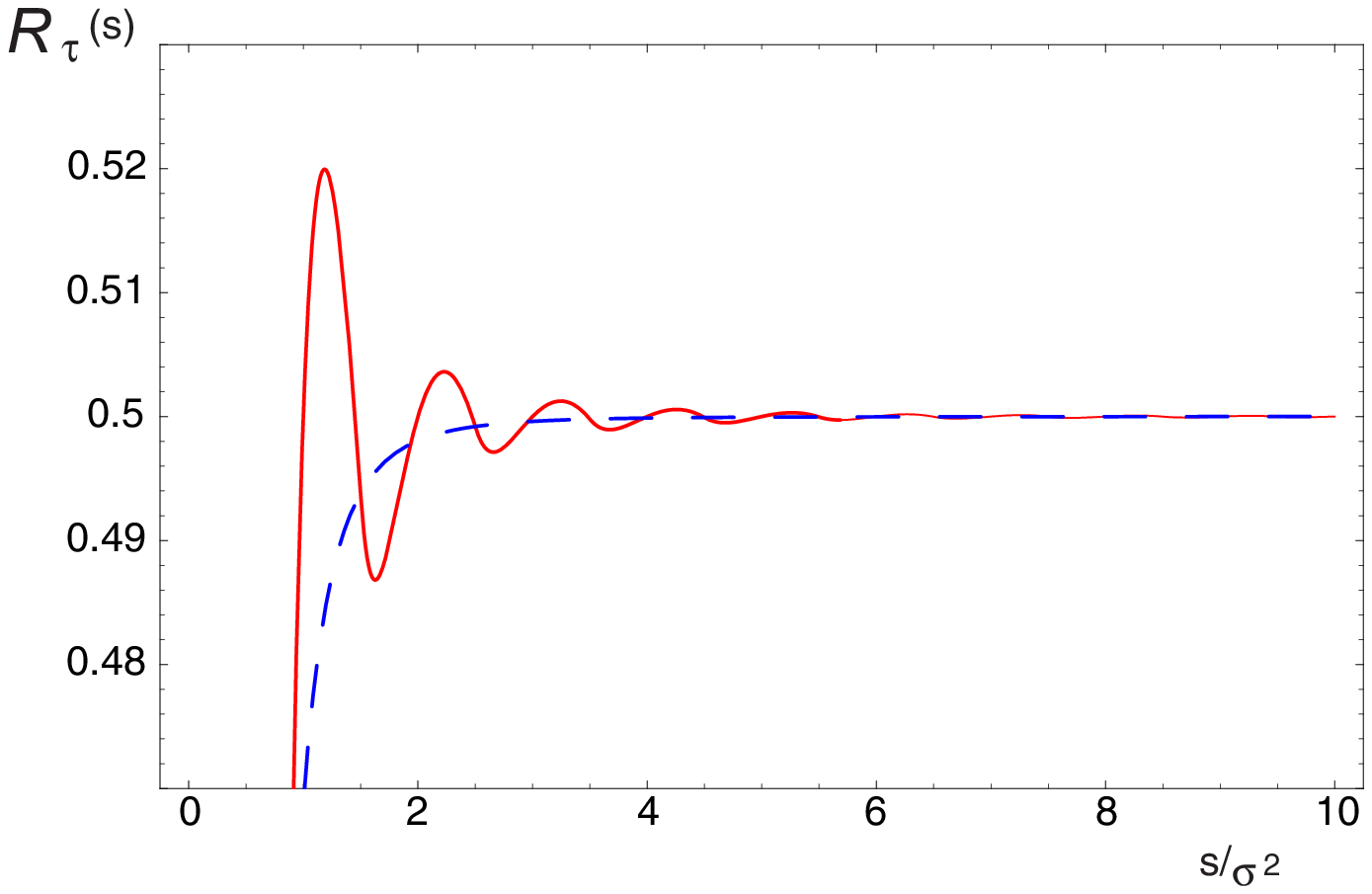}} \vskip 1pc {\bf
Fig.~6} {\it Plot of the branching ratio $\cR_{\tau}(s)$ defined
in Eq.~\rf{Rtau}. The continuous red curve is the one predicted
by the {\it exact} toy model; the dashed blue curve is the one
predicted using the OPE.} \vskip 2pc

However, one should also notice in this figure that there are
``optimal" points at $s/\sigma^2\approx m/2, m=2,3,4,\ldots$
where the difference between the oscillation and the OPE curve
nearly vanishes. These ``optimal" points are very useful because
they allow the OPE to produce a very accurate result even at very
low values of $s/\sigma^2$ ! We shall now elaborate on these
special points and explain where they come from. To this end it
is useful to identify clearly from $\mathcal{R}_{\tau}$ the
combination which goes into $\mathcal{R}_{\tau}^{\mathrm{OPE}}$
and the one which goes into $\mathcal{R}_{\tau}^{\mathrm{OSC}}$.
This can be readily obtained by rewriting $\cR_{\tau}(s)$ in
Eq.~\rf{Rexact} as a linear combination of Bernoulli polynomials
in the $x$ variable. Using Eqs.~(A.8) and (A.9) we can rewrite \be
\cR_{\tau}(s)=\cR_{\tau}^{\mathrm{OPE}}(s)+
\cR_{\tau}^{\mathrm{OSC}}(s)\,, \ee with
$\cR_{\tau}^{\mathrm{OPE}}(s)$ as given in Eq.~\rf{Rope}, and \be
\cR_{\tau}^{\mathrm{OSC}}(s)=-\left(\frac{\sigma^2}{s} \right)^3
B_{3}(x)+\frac{1}{2}\left(\frac{\sigma^2}{s} \right)^4
B_{4}(x)\,. \ee  The oscillatory nature of this expression follows
from the Fourier decomposition of the Bernoulli polynomials, as
given in Eqs.~(A.15) and (A.16), with the result \be\lbl{fourier}
\cR_{\tau}^{\mathrm{OSC}}(s)=- 12
   \sum_{n=1}^{\infty}\left\{\left(\frac{\sigma^2}{s}\right)^3
   \frac{\sin\left[2\pi n \left(\frac{s}{\sigma^2}-\frac{1}{2}\right)\right]}
   {\left(2\pi n\right)^3}+
   2\left(\frac{\sigma^2}{s}\right)^4
   \frac{\cos\left[2\pi n \left(\frac{s}{\sigma^2}-\frac{1}{2}\right)\right]}
   {\left(2\pi n\right)^4}
   \right\}\,,
\ee where we have used the fact that $x$, defined in Eq. \rf{x},
is equivalent to $\frac{s}{\sigma^2}-\frac{1}{2}$ in the argument
of $\sin\left[2\pi n x\right]$ and/or $\cos\left[2\pi n
x\right]$. Equation \rf{fourier} shows the oscillations
decomposed in terms of harmonics. Interestingly, one sees that
the leading oscillation at large $s$, i.e. $\sim 1/s^3$, cancels
exactly for values of $s/\sigma^2$ which are integers or
half-integers, i.e. $\frac{s}{\sigma^2}=\frac{k}{2}\,,$
$k=0,1,2,\dots$~\footnote{Notice that for the value $x=1/2$, the
parton model and the exact result in Eq.~\rf{Rexact} agree with
each other. However, this is no longer true when $\hat{M_0}^2
\neq 1/2$. See also footnote 5. } The set of half-integers
coincide with the exact location of the resonances and would
correspond in the case of true QCD to the masses in the
large-$N_c$ limit, which is information one does not have in real
life. However, the integer values coincide with the condition on
$s$ in Eq.~\rf{cont}, namely $s=(m+1)\sigma^2$. This is more
useful since it only relies on the matching of the MHA onto the
first term of the OPE which is always available.  Of course, it
is not known to what extent a similar relation there exists
between the leading oscillatory term and the $1/Q^2$ term in the
OPE for the case of large-$N_c$ QCD; these features may very well
be generic, and should be kept in mind when dealing with
phenomenological studies of finite--energy sum rules in QCD.
Indeed, the analysis of Ref. \cite{PPdeR01} shows that the data
seem to support this.

The representation in Eq.~\rf{fourier} neatly disentangles the
effect of the spectrum (the infinite sum over the oscillatory
terms) from its projection on the particular observable we are
considering; in this case $\cR_{\tau}$, which is only sensitive
to the two powers $\left(\sigma^2/s\right)^3$ and
$\left(\sigma^2/s\right)^4$.

Looking at Fig. 6 makes it clear that an analysis of the data at a
given value of $s$  based solely on the OPE curve induces an
error which is proportional to the size of the nearby oscillation.
Consequently it is highly desirable to develop an approximation
scheme which, at least in principle, can reproduce these
oscillations. We shall now see that the MHA is such a scheme.

One example of the prediction from the MHA for
$\mathcal{R}_{\tau}$ can be seen in Fig. 7. This example is the
case of four resonances which are matched onto the first two
long-distance and the first two short-distance constraints. Since
the MHA has been constructed by matching on to the first few terms
of the OPE, it has to reproduce the corresponding prediction of
the OPE for $\mathcal{R}_{\tau}$ as soon as $s>s_0$, $s_0$ being
the onset of the continuum of the MHA. Since $\mathcal{R}_{\tau}$
can only read the inverse powers up to and including $1/Q^8$,
this means the MHA will reproduce the result of
$\mathcal{R}_{\tau}^\mathrm{OPE}$ provided at least three
resonance masses are matched on to the corresponding three
short-distance constraints corresponding to the powers
$1/Q^4,1/Q^6,1/Q^8$ of the OPE (recall that the $1/Q^2$ power is
always used to determine $s_0$). Since in the MHA curve in Fig. 7
only the first two inverse powers of $Q^2$  have been used, this
explains the small difference between the MHA and the OPE curves
in Fig. 7 at $s=s_0=4\sigma^2$. One can check that this difference
disappears when the three short-distance constraints coming from
the $1/Q^4,1/Q^6,1/Q^8$ powers of the OPE are used. This is not
surprising. It is also not surprising that the MHA fits so well
the low--energy region in Fig. 7, since the MHA has been matched
onto the first two chiral powers at low energies. What is more
interesting is how well the MHA also describes the intermediate
region of $s$ with all the oscillations. This result clearly goes
beyond the input.

\vskip 2pc
\centerline{\epsfbox{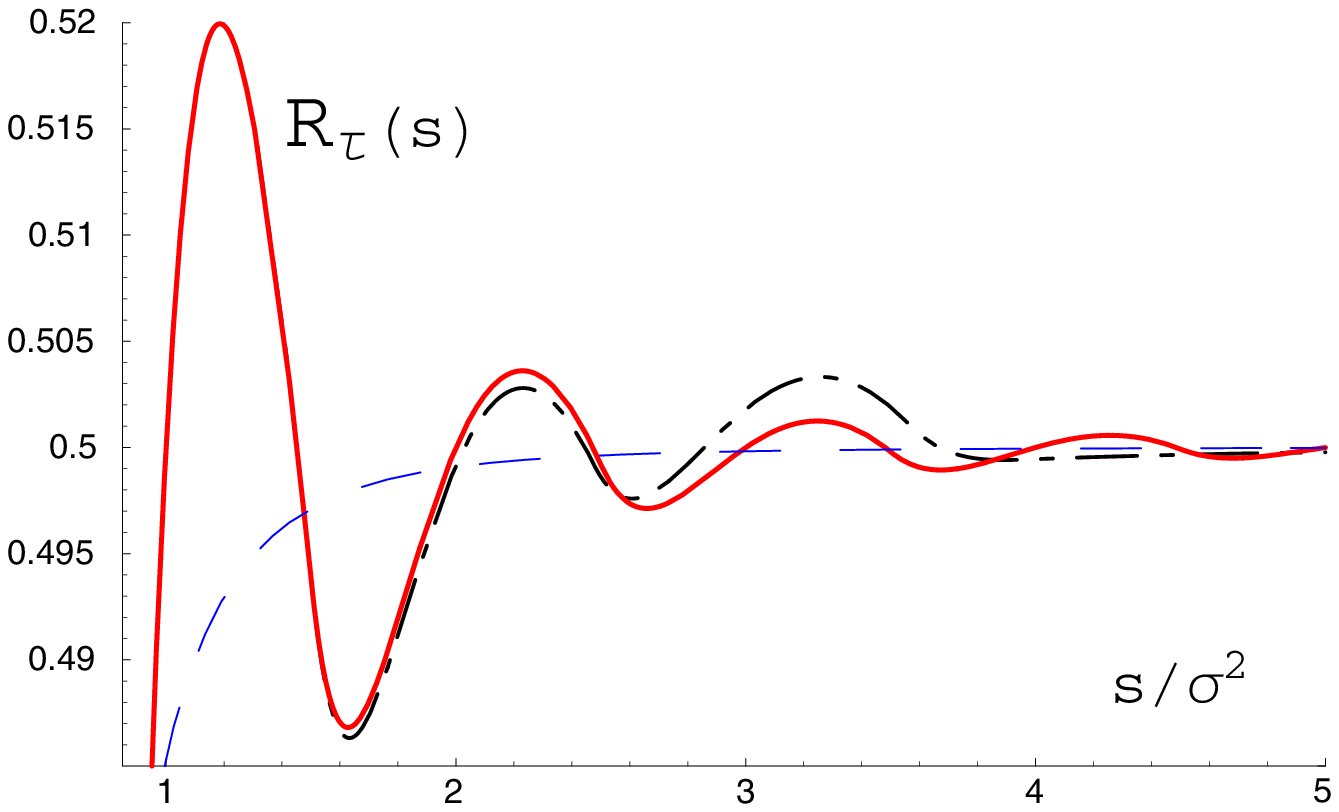}} \vskip 1pc {\bf Fig.~7}
{\it Plot of the branching ratio $\cR_{\tau}(s)$ defined in
Eq.~\rf{Rtau}. As in Fig. 5, the continuous red curve is the exact
result predicted by the toy model, the dashed blue curve is the
one predicted using the OPE. The black dot--dashed curve
is the prediction of the MHA.  This curve has been produced with
four resonances fixed by the first two short-distance and the first
two long-distance constraints.} \vskip 2pc

A useful conclusion, for phenomenological purposes,
which emerges from these toy--model exercises is that, when
confronting the real situation of a finite--energy sum rule
involving {\it hadronic data} to QCD, say a sum rule of the type
\be \int_{0}^{s_{0}} dt\;\rho(t)_{{\mbox{\rm\tiny hadronic}}}=
F_{{\mbox{\rm\tiny QCD}}}(s_{0})\,, \ee the $s_{0}$ values at
which the comparison between theory and data should be made are
not {\it arbitrary} \cite{PPdeR01}, as is unfortunately often
assumed in the literature; optimally, they must be made at the
specific {\it duality} values which are appropriate solutions of
the leading OPE constraints relevant to the case one is
considering. Alternatively, one can choose the highest value of
$s$ available by the data (which in the real world is fixed by
the $\tau$ mass) so as to minimize the size of the oscillations, and
then use the MHA to approximate their shape. It would be very
interesting to reanalyze $\tau$ data in these two new ways.
The case of QCD is of course more complicated than that of our
toy model, but our discussion of the toy model nevertheless
suggests that both ways of analyzing real data should prove
useful, and possibly lead to an estimate of the errors involved.
In addition, one can extend the use of the toy model to investigate
the MHA for other quantities of interest.  Finally, we recall
that a slightly more sophisticated version of the model
does remarkably well phenomenologically \cite{GP01}, indicating that it does
have something to do with the  real solution of large--$N_c$ QCD.

\section{\bf Acknowledgements}

E. de R. and S. P. would like to thank the organizers of the
workshop ``Lattice QCD and Hadron Phenomenology'' held at the INT,
University of Washington at Seattle, where this work was
finalized, for their kind invitation and warm hospitality. S.P.
would also like to thank J. Carmona and E. Elizalde
for discussions. M.G. thanks
the IFAE at the Universitat Aut\`onoma de Barcelona for
hospitality.

The work of E. de R.  and S.P. is supported by TMR, EC-Contract
No. ERBFMRX-CT980169 (EURODA$\phi$NE). The work of S.P. is also
supported by CICYT-AEN99-0766, and that of M.G. in part by the
US Dept. of Energy.

\vspace*{2cm}
\begin{center}
\appendix
\section{{\bf APPENDIX}}
\end{center}
\setcounter{equation}{0}
\def\theequation{\Alph{section}.\arabic{equation}}

\vspace*{0.5cm}
\noi
{\bf 1.~Bernoulli Polynomial}
\vspace*{0.25cm}

\noi
The Bernoulli polynomials are defined as:
\be
B_{n}(x)=\sum_{r=0}^{n}\left(\begin{array}{c} n \\ r \end{array}
\right) B_{r}\,\, x^{n-r}\,,
\ee
where
\be
\left(\begin{array}{c} n \\ r \end{array}
\right)=\frac{\Gamma(n+1)}{\Gamma(r+1)\Gamma(n-r+1)}\,,
\ee
and $B_{2n}$ are the Bernoulli numbers :
\be
B_{0}=1\,,\quad B_{1}=-\frac{1}{2}\,,\quad B_{2}=\frac{1}{6}\,,\quad
 B_{4}=-\frac{1}{30}\,,\quad
B_{6}=\frac{1}{42}\,,\quad\cdots\,;
\ee
with
\be
B_{2n+1}=0\,,\qquad\foor\qquad n\ge 1\,.
\ee
Examples of the first few Bernoulli polynomial are:
{\setl
\bea
B_{0}(x) & = & 1 \,, \\
B_{1}(x) & = & x-\frac{1}{2}\,, \\
B_{2}(x) & = & x^2-x+\frac{1}{6}\,, \\
B_{3}(x) & = & x^3 -\frac{3}{2}x^2+\frac{1}{2}x\,, \\
B_{4}(x) & = & x^4-2x^3+x^2-\frac{1}{30}\,.
\eea}

\noi
The Bernoulli numbers are related to the
Riemann zeta function of even argument:
\be
B_{2n}=(-1)^{n+1}(2\pi)^{-2n}2(2n)!\zeta(2n)\,,\quad\foor\quad
n=0,1,2,\dots\,.
\ee
Other important relations for Bernoulli polynomials are:
{\setl
\bea
B_{n}(x=0) & = & B_{n}\,, \\
B_{n}(x+1)-B_{n}(x) & = & nx^{n-1}\,, \\
\lbl{half} B_{n}(1-x) & = & (-1)^{n}B_{n}(x)\,, \\
B'_{n}(x) & = & nB_{n-1}(x)\,.
\eea}
It is sometimes useful to recall the trigonometric expansion of
Bernoulli polynomials, for $N\ge 1$:

{\setl
\bea\lbl{bertrigoeven}
B_{2N}(x) & = & (-1)^{N+1}
2(2N)!\frac{1}{(2\pi)^{2N}}\sum_{n=1}^{\infty}\frac{\cos 2\pi
nx}{n^{2N}}\,,
\\\lbl{bertrigoodd} B_{2N+1}(x) & = & (-1)^{N+1}
2(2N+1)!\frac{1}{(2\pi)^{2N+1}}\sum_{n=1}^{\infty}\frac{\sin 2\pi
nx}{n^{2N+1}}\,.
\eea}


\end{document}